\documentclass[12pt ,onecolumn]{IEEEtran}
\usepackage{times}
\usepackage{amssymb}
\usepackage{amsmath}
\usepackage{graphicx}
\usepackage[utf8]{inputenc}
\usepackage{setspace}
\usepackage{psfrag}
\usepackage{cite}
\usepackage{cleveref}
\usepackage{siunitx}
\usepackage{pstricks}

\doublespace

\usepackage[nolists,tablesfirst,nomarkers]{endfloat}

\begin{document}

\title{Dynamics of Complex Systems Built as Coupled Physical, Communication and Decision Layers}
\author{Florian Kühnlenz, Pedro H. J. Nardelli}

\maketitle
\markboth{Submission  v1: \today}{}

\begin{abstract}
This paper proposes a simple model to capture the complexity of multi-layer systems where their constituent layers affect, are affected by, each other.
The physical layer is a circuit composed by a power source and resistors in parallel.
Individual agents can add, remove or keep the resistors they have, and their decisions aiming at maximising the delivered power -- a non-linear function dependent on the others' behaviour -- based on their internal state, their global state perception, the information received from their neighbours in the communication network, and a randomised selfishness.
We develop an agent-based simulation to analyse the effects of number of agents (size of the system), communication network topology, communication errors and the minimum power gain that triggers a behavioural change.
Our results show that a wave-like behaviour at macro-level (caused by individual changes in the decision layer) can only emerge for a specific system size, the ratio between cooperators and defectors depends on minimum gain assumed -- lower minimal gains lead to less cooperation and vice-versa, different communication network topologies lead to different levels of power utilisation and fairness at the physical layer, and a certain level of error in the communication layer leads to more cooperation.
\end{abstract}

%
\section {Introduction}

The modernisation of large-scale engineering infrastructures brings to the table new challenges to their already complicated design \cite{helbing2013globally,nikolic2009co,van2012agent}.
For example, electric power grids are large-scale engineering systems built to generate, transmit and distribute electricity from generators to end-users \cite{kremers2013modelling,nardelli2014models,bush2014smart}.
Although their technological development has never stopped, a strong political demand for a structural change is taking place.
Such change basically consists in decentralising generating units (e.g. from nuclear to solar panels and wind turbines), spreading of electric vehicles (which are mobile batteries and loads) and controlling demand based on information technologies; all in all, the traditional consumer is predicted to become a \textit{prosumer}: a consumer who participate more actively in the grid management either by supplying electricity or decreasing their consumption.
Modern power grids will become more dynamic and distributed.
this will bring new complexities to their already complex dynamic together with new research challenges to cope with them.

The same trends -- although in with their own specificities --  can be seen when analysing the modernisation of other large-scales systems, from smart cities\cite{batty2007cities,batty2013cities} to factories of the future \cite{herrmann2014sustainability} or the 5-th generation of cellular systems \cite{hossain2014evolution}.
As in power grids, new complexities in those systems will emerge followed by a need for a new body of knowledge.
Notwithstanding the unquestionable technical evolution, there is still a limited number of simple analytic models that are able to capture the dynamics of these modern systems, where the physical infrastructure, the information network and regulations affect, and are affected, by each other dynamics.
In this context the present article proposes a discrete-time agent-based model assuming these three layers as constitutive parts of a multilayer system composed by an electric circuit as the physical infrastructure, a communication network where agents exchange local information and a set of regulations that define the agents' behaviour.
\begin{figure}[!t]
\centering
\includegraphics[width=0.8\columnwidth]{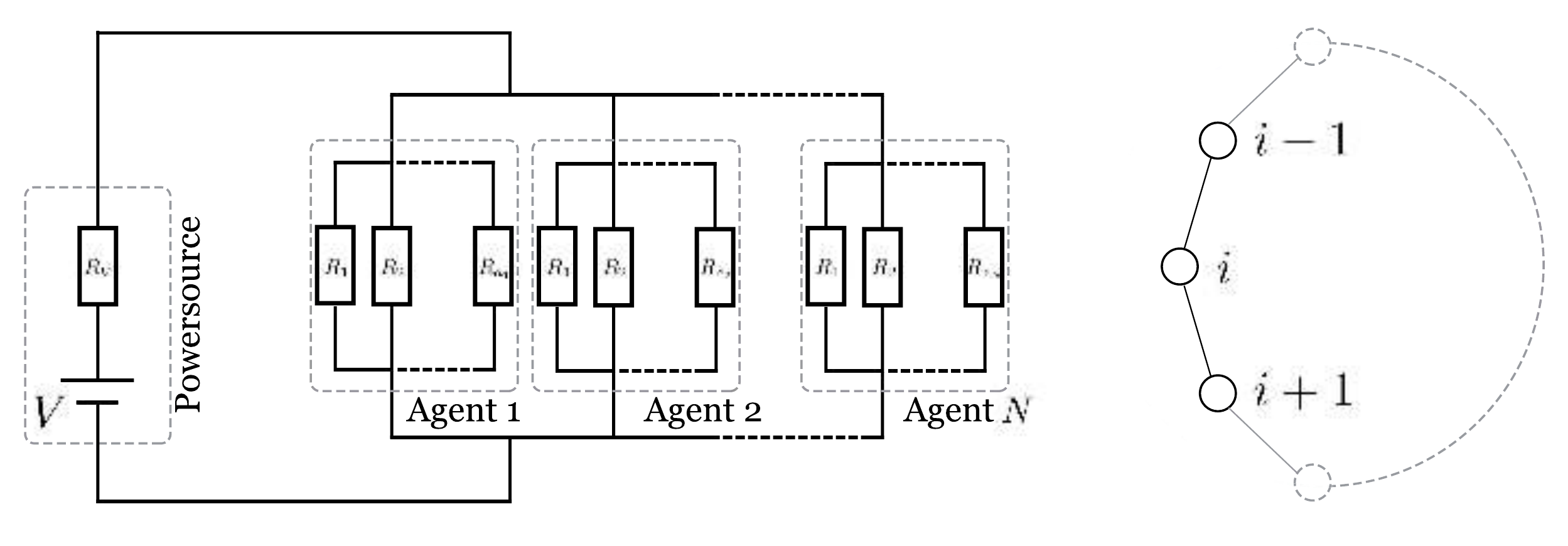}
\caption{Electrical circuit representing the physical layer of the system. The circuit is composed by a power source ($V$ and $R_\mathrm{V}$, and resistors of $R$ in parallel. These resistors are related to $N$ agents that can add, remove or keep the resistors under their control in the circuit.
The minimum number of resistors an agent can have is one and there is no maximum.
We also consider $N$ as the size of the system.
Besides the physical layer, the agents are connected in a communication network so that a given agent has access to the information related to the previous action of their first-order neighbours \cite{Newman2006}. In the ring topology illustrated here, every agent is connected with two other agents. In this case agent $i$ is connected with agents $i-1$ and $i+1$ with $i = 1,2,...,N$. In the ring topology, agents $1$ and $N$ are neighbours. \label{fig:Electrical-System}}
\end{figure}

\begin{figure}[!t]
\centering
\includegraphics[width=0.5\columnwidth]{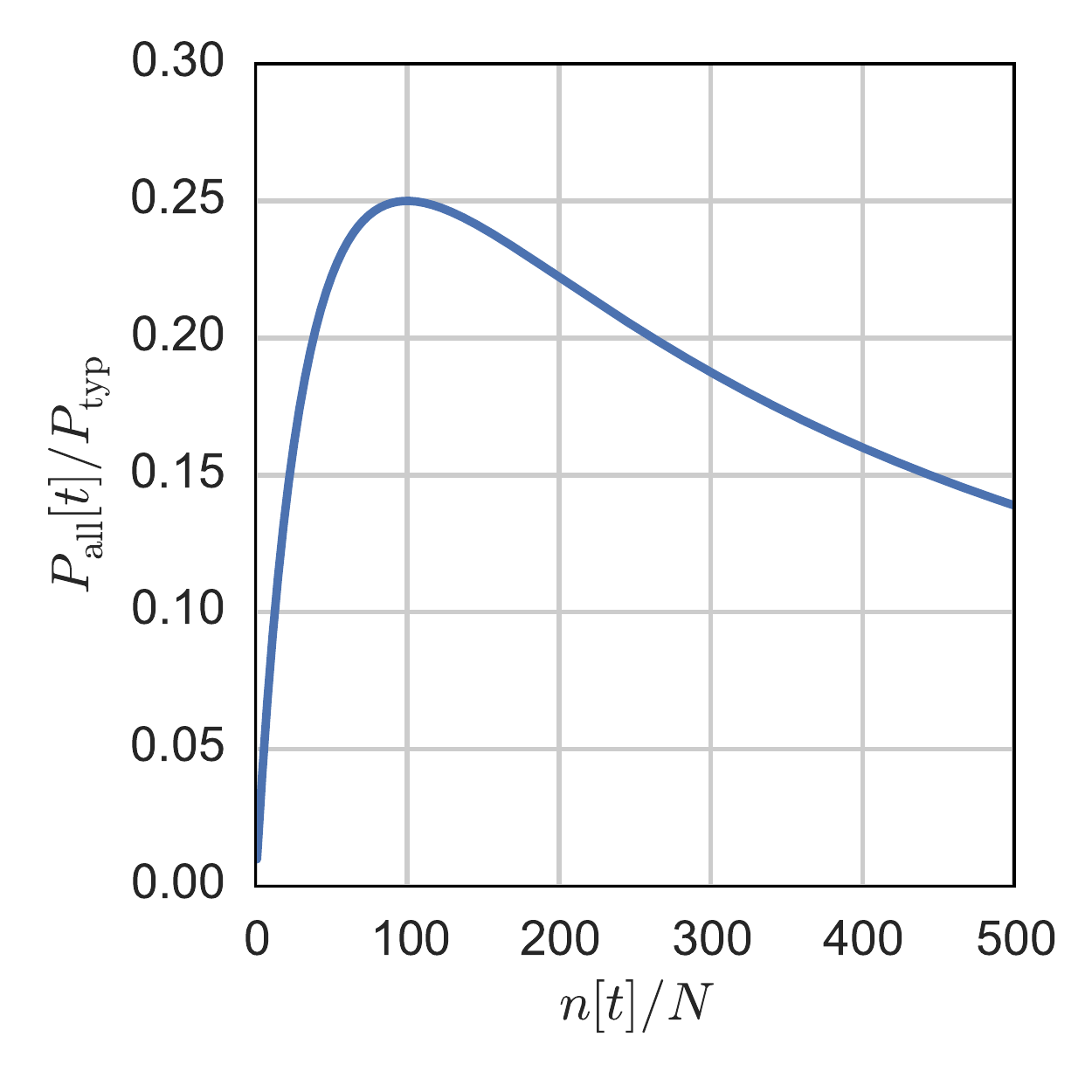}
\caption{\label{fig:Power-System-Behaviour} Normalised power delivered $P_\mathrm{all}[t]/P_\mathrm{typ}$ to the agents with rising number of resistors $n[t]$ in the system, where $P_{\mathrm{typ}}=\frac{V^{2}}{R_\mathrm{V}}$ and $P_\mathrm{all}[t] = \sum_{i\in \mathcal{A}} P_i[t]$ with $P_i[t]$ given by (1).}
\end{figure}

The proposed model is built as follows.
The electric circuit is composed by one constant voltage source including its inner resistance and resistors (loads) in parallel \cite{dorf2010introduction}.
The resistors in parallel are grouped by their controlling agent, as shown in Figure \ref{fig:Electrical-System}.
Every agent may switch on or off one of the resistors under its control at every time step.
One can expect that, the greater the number of active resistors a given agent has, the more power is delivered to it.
The actual delivered power is, however, a non-linear, concave, function (as presented in Figure \ref{fig:Power-System-Behaviour}) of the electric current flowing in the circuit; there exists then a saturation point where adding more resistors will decrease the delivered power for the whole system.

Looking at the whole system, a ``tragedy of the commons'' kind of problem arises \cite{Hardin1968}, where add a resistor is individually beneficial, while socially harmful.
However, in our case, the resource recovers very quickly.
The agent's decision regarding the resistors (adding, subtracting or maintaining) is built upon the following criteria: the behaviour of the agent's neighbours at the last time step, the previous state of the whole system and its own selfishness gene.
As we will discuss later, this interactive decision procedure resembles the prisoners dilemma (e.g \cite{szabo2007evolutionary, archetti2012review,Gianetto2015} and references therein)
In this case the agents' neighbourhood is defined by the communication network (where links can be in error), and the selfishness genes of the agents are independent and identically distributed random variables.

\subsection{Complexity sciences}
Before we start presenting our contribution in more details, it is worth indicating the theoretical ground  that supports our findings: complexity sciences.
Complexity is a term used in several diverse research fields \cite{Mitchell2009,Furtado2014}, from theoretic physics to social sciences and biology, to characterise a state that is neither completely deterministic nor random.
The so-called complex behaviour emerges in systems whose elements interact; they may be heterogeneous and may also adapt their relation rules in accordance to internal and/or external factors.
In its extensive work \cite{Wolfram2002}, Wolfram has shown that simple interaction rules applied in one-dimensional cellular automata may lead to unexpected intricate patterns  -- defined therein as complex -- over time.
this work tells us that, when looked at a higher level, the spatial-temporal dynamics of a fairly simple deterministic system composed by homogeneous agents that follow fixed interrelation rules may generate complexity.
This fact suggests that decentralised systems, based only in local information, might be functional without any controlling entity.
In Wolfram's case the spatial-temporal pattern is determined by the interaction rule\footnote{Following its classification, four kind of behaviours can emerge in the one-dimensional automata, namely homogeneous, periodic, chaotic and complex. Refer to \cite{Wolfram2002,Furtado2014} for more details.}.

For some researchers, this fact indicates the system is able to self-organise without any explicit centralised controller.
We can cite here few illustrative examples regarded as self-organised \cite{Mitchell2009}: ants working in colonies, neurons building a capable brain and birds flying in groups.
As an interesting counter point to this perspective, one may argue that the interaction rule that the agents follow is \textit{per se} a kind of central control or a strict regulative force \cite{hodgson2012pleasure}.
By using this view, many questions may be posed: From where the interaction rules comes from? Are they evolving? Are they changing in a much slower time-scale that can be considered as given?
These questions in fact still cause hot philosophical debates among biologists, economists, social scientists and other theorists concerning who controls the ``invisible hand"; more details about these debates can be found in, for example,  \cite{hodgson1997economics} and references therein.

When dealing with large-scale engineering systems (the focus of this paper), these questions seem to have clear answers as far as the system is designed and follows pre-defined requirements.
This is true for some systems, but it is far from being a universal feature, mainly when the infrastructure is heavily dependent on human actions and interactions.
Road networks provide an educative, well studied, example of this when one tries to understand the formation of traffic jams \cite{helbing2001traffic}.
Without going into further details, the key for solving the traffic puzzle is not found by looking at what happens in individual cars or in the design of the whole transportation infrastructure.
While these aspects are necessary conditions to the formation of the jam, they are not sufficient to explain the phenomenon.
The most accepted theory is built upon the interactions between cars and reactions to the individual behaviours within a specific region of the road network; one car slowing down in highly dense highway causes other nearby cars to slow down as well to avoid collisions, which in turn may trigger a traffic jam that will fade away after some time.

This simple example identifies few important characteristics of complex phenomena: they are spatial and temporal, they never reach stable equilibrium states, the context where the individuals interact and its perception are important to individual decisions, and one individual action might cause a local change that might also trigger changes in the global state of the system.
More details about the so-called complexity sciences can be found in, for instance, \cite[Ch.1]{arthur2014complexity}.

\subsection{Contributions}
Motivated by a growing literature in complexity sciences in general and their application in engineering systems in specific, this articles present a new perspective to analyse multilayer, strongly coupled, systems.
We construct a simple, while illustrative, multilayer model composed by agents that control a set of resistors in an electrical circuit.
These agents play an evolutionary ``prisoners' dilemma" style of game to decide if they should collaborate or not, based on the local information gathered from their communication network, the estimated state of the whole system and their own random selfishness.
Our results indicate that: (i) a wave-like behaviour at macro-level spatio-temporal dynamics, which is caused by  changes in individual behaviours at the decision layer, can only emerge for a specific system size, (ii) the ratio between cooperators and defectors depends on minimum gain assumed -- lower minimal gains lead to less cooperation and vice-versa, (iii) different communication network topologies -- ring, Watt-Strogatz-Graph and Barabasi-Albert-Graph \cite{Newman2006} -- lead to different levels of power utilisation and fairness at the physical layer, and (iv) a certain level of error in the communication layer leads to more cooperative behaviour at the decision layer, affecting the physical layer dynamics in terms of power utilization and fairness accordingly.

\section{Results}
In this section we present the main results of this report.
Before starting, we think it is worth describing here the agents' decision process, which is a key point for understanding our results.
We also systematise  in Table \ref{table:notations} some useful notation to facilitate our analyses.
More details about the multi-layer system proposed here can be found later in Section \ref{sec-methods} (Methods).
\begin{table}[t]
\vspace{2ex}
\renewcommand{\arraystretch}{1.}
\caption{Notations}
\label{table:notations} \centering
\begin{tabular}{l|l}

Notation & Meaning   \\
\hline
$t \in \mathbb{Z}$ & discrete time \\
$\mathcal{A} = \{1,2,...,N\}$ &  set of all agents, where $N$ is the size of the system \\
$i \in \mathcal{A}$ & agent $i$\\
$\mathcal{N}_i \subset \mathcal{A}$ &  neighbourhood set of agent $i$\\
$n[t] \in \{N,N+1,...\}$ &  active number of resistors in the system at time $t$\\
$a_i[t] \in \mathbb{N}^+$ &  active resistors of agent $i$\\
$r_i[t] = n[t] - a_i[t]$ &  active resistors excluding agent $i$\\
$P_i[t] > 0$ &  consumed power of agent $i$ [in units of power]\\
$\lambda_i[t] \in \mathbb{R}$ &  gain in power of agent $i$\\
$\lambda_\mathrm{min} \in \mathbb{R}$ &  system-wide pre-defined minimum gain\\
$S_i[t] \in \{-1,0,+1\}$ &  state of agent $i$\\
$s_i \in [0,1]$ &  selfishness gene of agent $i$\\
$p_{\mathrm{err}} \in [0,1]$ & error probability in a communication link\\
\end{tabular}
\end{table}

\subsection{Agents' decision process}

We assume a discrete-time system such that the changes in the agent behaviour occur in time-steps, denoted by $t \in \mathbb{Z}$.
At every time-step $t$, each agent wants to maximise its own power, so their interactions can be then viewed as a round-based game \cite{archetti2012review}.
To achieve that goal, the agent has three options: add a resistor, remove it, or do nothing.
Table \ref{table:actions} shows how we classify the agent behaviour.
\begin{table}[b]
\vspace{2ex}
\renewcommand{\arraystretch}{1.}
\caption{Classification of agent $i\in \mathcal{A}$ behaviour based on its action at time $t$}
\label{table:actions} \centering
\begin{tabular}{l|c|r}

Physical action & Behaviour class & $S_i[t]$  \\
\hline
add resistor & defect & $ +1 $\\
remove resistor & cooperate & $-1 $\\
do nothing & ignore & $0$\\

\end{tabular}
\end{table}

To make a decision at time $t$, every agent $i$ looks at what his gain from the previous strategy $S_i[t-1]$ was, as to decide its new state $S_i[t]$.
The decision process for the agent $i$ is the following.
If the gain $\lambda_i[t-1]$ is greater than or equal to a system-wide pre-defined minimum $\lambda_\mathrm{min}$, the agent sticks to its (successful) strategy at time $t$, i.e. $S_i[t] = S_i[t-1]$.
If $\lambda_i[t]<\lambda_{\mathrm{min}}$, then agent $i$ compares its  strategy with its neighbourhood $\mathcal{N}_i$, which will be defined later in this section.
If the majority is cooperative, then $\sum\limits_{j \in \mathcal{N}_i} S_j[t-1] < 0$ and the agent under analysis will also cooperate, leading to $S_i[t] = -1$.
Otherwise, the agent draws a random number between 0 and 1 to be compared to its own selfishness gene $s_i$ (which is also randomly generated as discussed later) in order to decide whether it will start cooperating.
If it does not cooperate, it again draws a random number to be compared to the selfishness gene $s_i$, but now to decide if stays inactive (i.e. $S_i[t] = 0$) or adds another load in the circuit (i.e. $S_i[t] = +1$).

\subsection{Communication network}
In the multi-layer system proposed here, agent $i$ knows the state $S_j[t-1]$ of the agents $j \in \mathcal{N}_i$ through a communication network.
We assume that agent $j$ always transmit its actual state $S_{j}[t]$ to agent $i$.
The neighbourhood $\mathcal{N}_i$ of agent $i$ is defined as the agents $j \in \mathcal{A} \backslash \{i\}$ that are directly linked with it.
In the case of ring topology, the cardinality of $\mathcal{N}_i$ is 2 for all agents $i \in \mathcal{A}$.
For more complex network topologies, $\mathcal{N}_i$ will be characterised differently  \cite{Newman2006}, as discussed later.

The communication links can also experience errors.
An error event means the received message by agent $i$ contains a different information than agent $j$ has sent.
Let $S_{j\rightarrow i}[t-1] = S_{j}[t-1]$ be the state information transmitted from $j$ to $i$ at time-step $t$ and $\hat{S}_{j\rightarrow i}[t-1]$ be the information received by $i$.
We consider that error events are independent and identically distributed such that $\mathrm{Pr}\left[\hat{S}_{j\rightarrow i}[t-1] \neq S_{j\rightarrow i}[t-1] \right] = p_\mathrm{err}$ for all $t \in \mathbb{Z}$, $i \in \mathcal{A}$ and $j \in \mathcal{N}_i$.
It is worth mentioning that the network is a bidirectional graph so that an error event at $i \rightarrow j$ does not imply an error event at $j \rightarrow i$, and vice-versa.

If an error event happens, the received information $\hat{S}_{j\rightarrow i}[t-1]$ will be uniformly and identically distributed between the other possible states.
For example, if $S_{j\rightarrow i}[t-1] = -1$ and error happens, then $\hat{S}_{j\rightarrow i}[t-1] = 0$ or $\hat{S}_{j\rightarrow i}[t-1] = +1$ will happen with 50\% chance each.

\subsection{Physical system}
For the physical systems presented in Figure \ref{fig:Electrical-System}, there exists a certain number of resistors that leads to the maximum power gain in the system, as shown at Figure \ref{fig:Power-System-Behaviour}.
If the delivered power is below the maximum on the right, then there will be a gain by removing a resistor until the system has reached such point.
Conversely, if it is below on the left, then there will be a gain by adding a resistor.

In this case, one may ask the following question: \textit{Is it possible to reach the optimal point while fairly delivering power among the agents (i.e. they consume about the same amount of power)?}
In the presence of a central control unit this would be a fairly easy problem: first finding the number of resistors that lead to optimal power, and then fairly allocating them among the agents by some kind of centralised coordination mechanism like time division in computer networks or cellular systems \cite{marsic2010computer}.
For example, if there are ten agents and the optimal number of resitors is twenty, then the central control coordinates the behaviour such that all ten agents have two active resistors, summing up twenty.
However, as discussed before, our model does not consider the presence of a central control and the agents have a limited knowledge about other agents.

At time-step $t$, the power each agent consumes $P_i[t]$ with $i\in\mathcal{A} = \{1...N\}$ and is given by:

\begin{equation}
	\label{eq:power_i}P_{i}[t]=P_{\mathrm{typ}}\frac{a_i[t]\mu}{(a_{\mathrm{avg}}[t]+\mu)^{2}},
\end{equation}
where $P_{\mathrm{typ}}=\frac{V^{2}}{R_\mathrm{V}}$, $\mu=\frac{R}{R_{\mathrm{V}}}$, $a_i[t]$ is the number of active resistors the agent $i$ possesses, $r_i[t]$ is the number of active resitors in the system excluding the source resistor $R_\mathrm{V}$ and the ones controlled by agent $i$, and $a_{\mathrm{avg}}[t]=(a_i[t]+r_i[t])/N$.
The physical system is then described by its size $N$, the ratio $\mu$ of the resistance values and the power source $V$.
The resistors are scaled so that the optimal average number of resistors $\left(a_{\mathrm{avg}}^{*}\right)$ is independent of $N$ while the voltage might be scaled with $\sqrt{N}$ to have a constant ratio of power per agent, as explained in Section \ref{sec-methods}.
The gain that agent $i$ experiences at time-step $t$ is then defined as:

\begin{equation} \label{eq:gain_i}
	\lambda_i[t]=\frac{P_{i}[t]-P_{i}[t-1]}{P_{i}[t-1]} = \dfrac{\Delta P_i}{P_{i}[t-1]}.
\end{equation}

This implies that the agents only use the information about the previous time-step $t-1$.
If we expand (2)  using (1), the equation that determines $\lambda_i[t]$ becomes more complicated.
To make the analysis clearer, we choose to apply the following approximation (more details in Section \ref{sec-methods}):

\begin{equation}
\lambda_i[t] \approx \frac{\mathrm{d} P_{i}}{P_{i}[t]}\approx\frac{\Delta a_i[t]}{a_i[t]}-\frac{\;2\;}{N} \; \frac{1}{a_{\mathrm{avg}}[t]+\mu}\left(\Delta r_i[t]+\Delta a_i[t]\right),
\label{eq:gain}
\end{equation}
such that the gain $\lambda_i[t]$ is now a function of the variations in agent $i$'s own number of resistors $\Delta a_i[t] = a_i[t]  - a_i[t-1]$ and in the number of resistors controlled by other agents $\Delta r_i[t] = r_i[t] - r_i[t-1]$,  as well as the average number of resistors $a_{\mathrm{avg}}[t]$ and the system parameters $N$ and $\mu$.

\begin{figure}[!t]
\centering
\includegraphics[width=1\columnwidth]{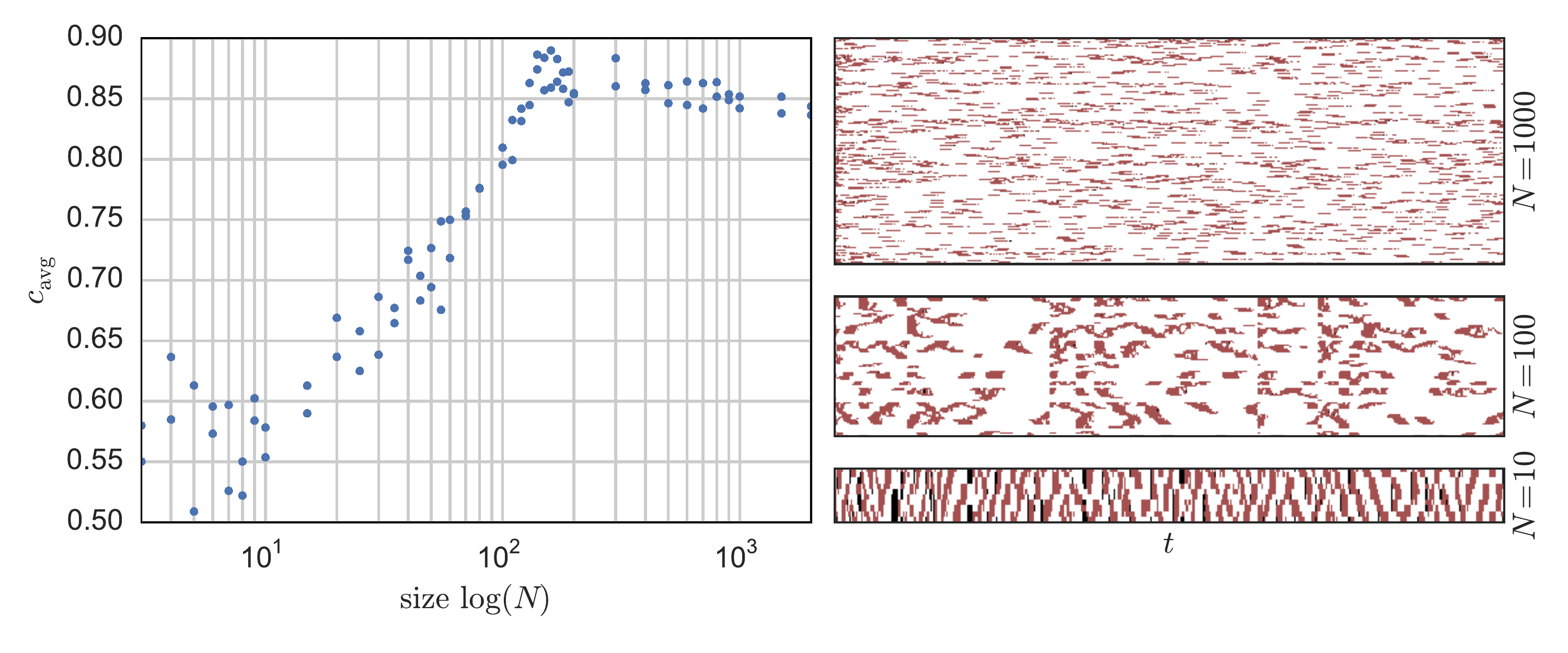}
\caption{\label{fig:Phase-Transition}On the left: change in the average cooperation $c_\mathrm{avg}$
depending on the system size $N$. On the right: examples of a typical system
behaviour as a function of time $t$ where the points (pixels) represent the agent state $S_i[t]$ such that red means defection, white cooperation, and black 
doing nothing for $N=1000$ (top), $N=100$ (middle) and $N=10$ (bottom). The system parameters are: $\lambda_\mathrm{min}=0.0005$, $R_V=\SI{2}{\ohm}$, $R_0= R/N = \SI{200}{\ohm}$, $p_\mathrm{err}=0.01$ and $V=\SI{1}{V}$.}
\end{figure}

\begin{figure}[!t]
\centering
\includegraphics[width=1\columnwidth]{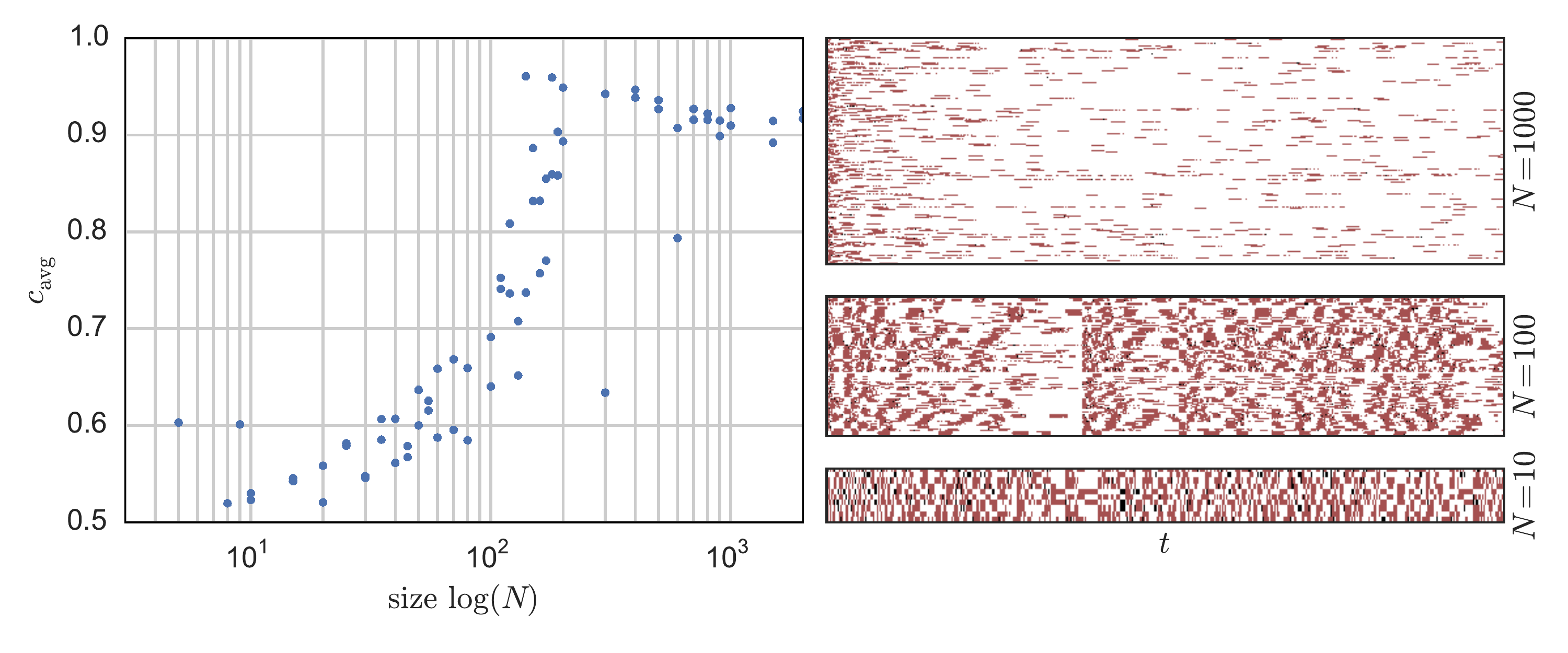}
\caption{\label{fig:transition-WS} Communication network based on Watts-Strogatz graph \cite{Newman2006}. On the left: change
in the average cooperation $c_\mathrm{avg}$ depending on the system size $N$. On the right: examples of a typical system
behaviour as a function of time $t$ where the points (pixels) represent the agent state $S_i[t]$ such that red means defection, white cooperation, and black =
doing nothing for $N=1000$ (top), $N=100$ (middle) and $N=10$ (bottom). The system parameters are: $\lambda_\mathrm{min}=0.0005$, $R_V=\SI{2}{\ohm}$, $R_0= R/N =\SI{200}{\ohm}$, $p_\mathrm{err}=0.01$ and $V=\SI{1}{\volt}$.}
\end{figure}

Let $\mathcal{C}_t \subseteq \mathcal{A}$ be the cooperative agents at time-slot $t$.
In this case, $\#\left(\mathcal{C}_t\right) \leq \#\left(\mathcal{A}\right) = N$ where $\#\left(\cdot\right)$ is the cardinality of the set.
The spatial-temporal average number of cooperators in the system ($c_\mathrm{avg}$) is then:
\begin{equation}
c_\mathrm{avg} = \dfrac{1}{N} \; \lim\limits_{T \rightarrow \infty} \; \dfrac{1}{T} \sum\limits_{t=0}^{T-1} \#\left(\mathcal{C}_t\right).
\end{equation}

Figure \ref{fig:Phase-Transition} shows the change of the system behaviour  with varying size $N$.
The left side shows the changes in the spatial-temporal average number of cooperators in the system $c_\mathrm{avg}$ as a function of the system size $N$, while the right side shows a representation of a typical spatial-temporal system behaviour for three different sizes $N$.
In the latter, each line of vertical pixels represents the state of the system for one time-step: white means cooperation, red means defection and black means doing nothing.

For small sizes (as when $N=10$), one can see a kind of checkerboard pattern where cooperators and defectors alternate on time and space axes.
For middle-sized systems (as when $N=100$), the most striking feature is that there exist certain points in time when sudden changes in the behaviour happen.
The system seems to move closer and closer to a global cooperation state until suddenly it falls back to a state with much less cooperation.
Such pattern becomes even more pronounced for a meshed communication network built as a Watts-Strogatz graph \cite{Newman2006}, as seen in Figure \ref{fig:transition-WS}.
For larger systems (as when $N=1000$), one can see that a pattern where cooperation is dominant and only few stripes of defecting appear.
As we will see next, this behaviour is not due to simple scaling effects in the system variables, but rather it results from inherent scaling effects within the physical layer. 
In other words, certain behaviours can be only observed for certain system sizes $N$.
Therefore, for a given $N$, one cannot pre-set the system variables expecting a certain kind of behaviour. 
Rather, the size of the system is itself a variable that influences its on macro-level behaviour.

\begin{figure}[!t]
\includegraphics[width=0.5\columnwidth]{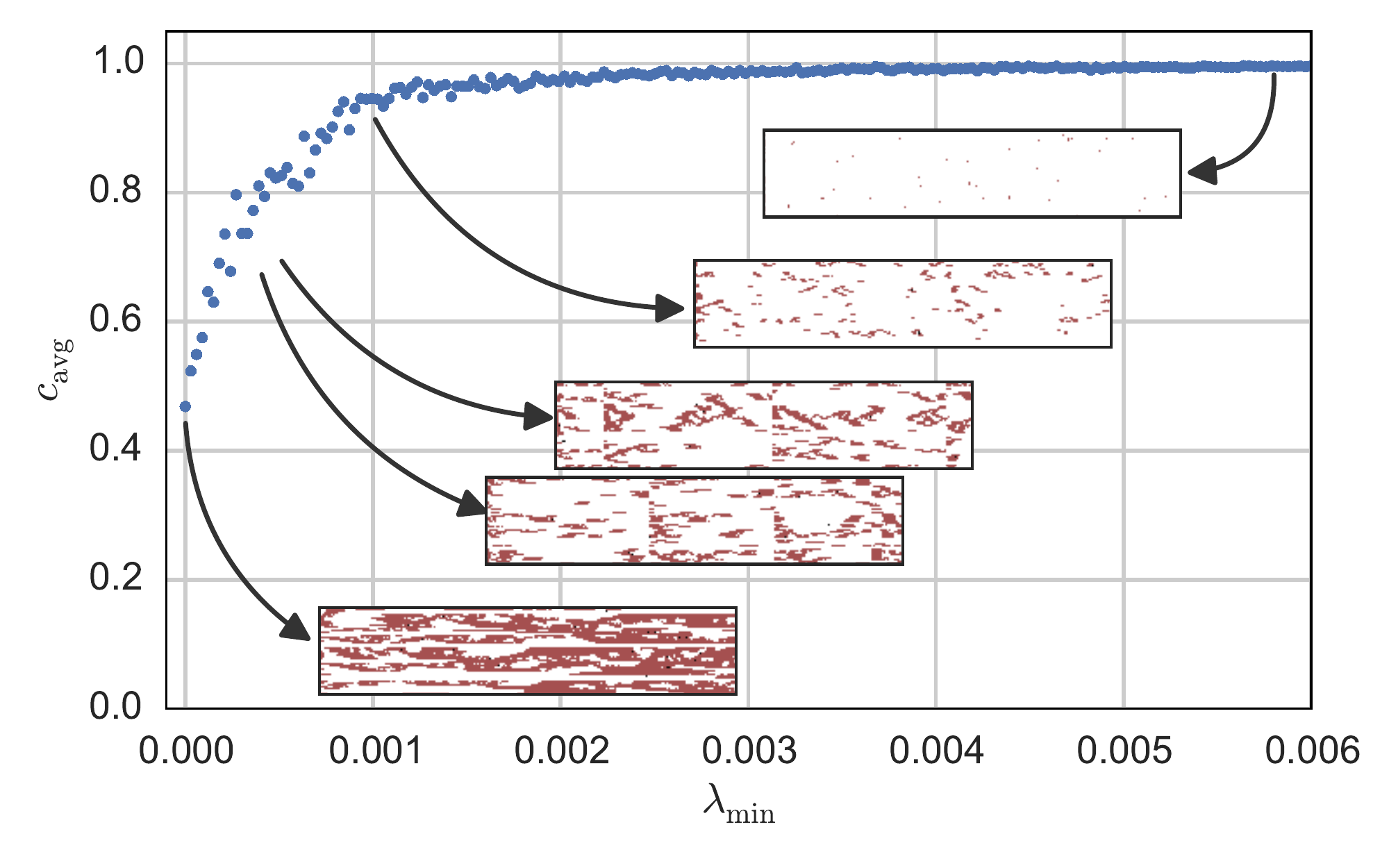}
\includegraphics[width=0.5\columnwidth]{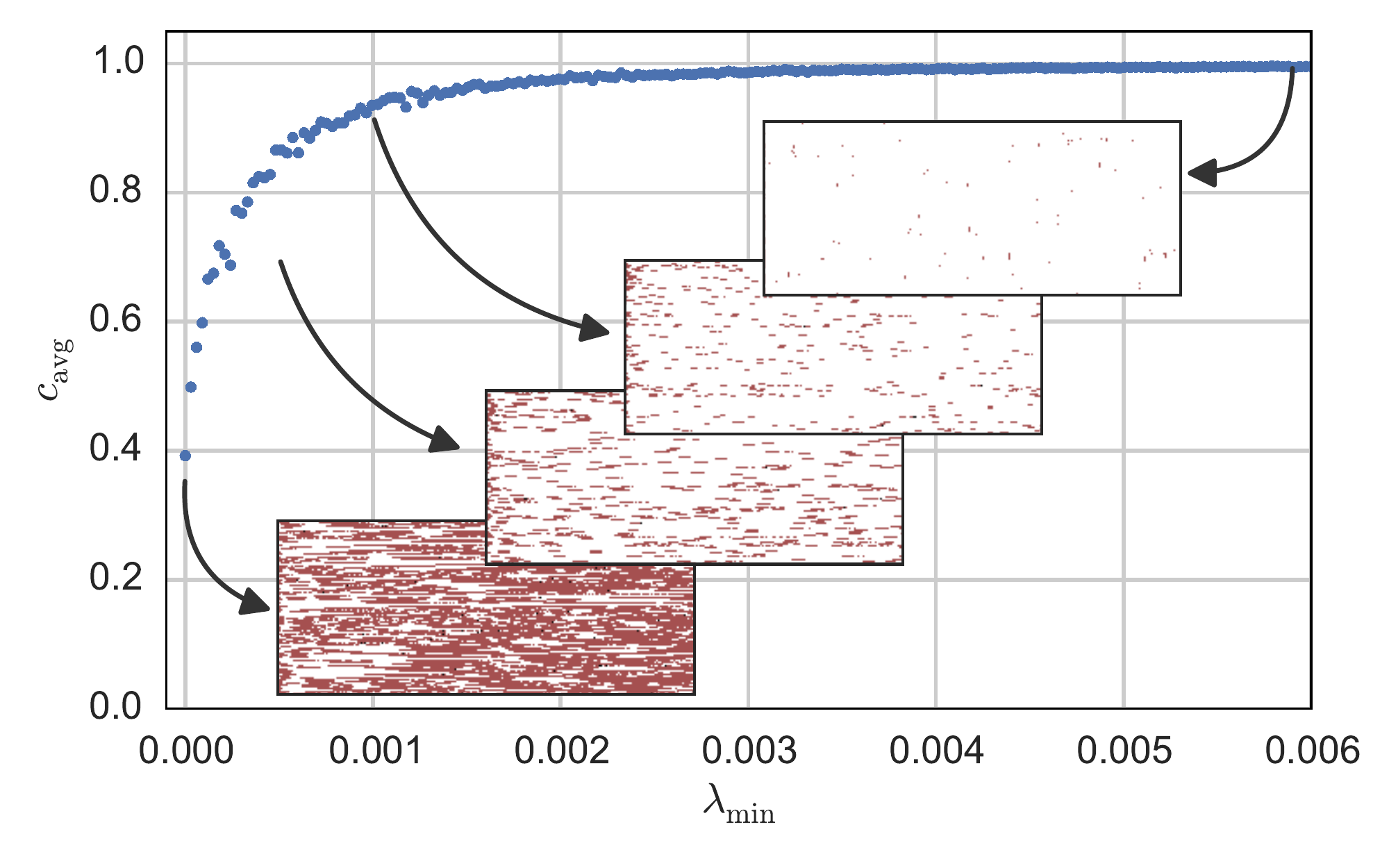}
\caption{\label{fig:minimal-gain} Influence of minimal gain $\lambda_\mathrm{min}$ on system behaviour for $N=100$ (left) and $N=500$ (right). Bigger minimum gains leads to more cooperation while smaller ones lead to defection. More complex behaviour can be observed for values close to $\lambda_{\mathrm{min}}=0.0001$ in mid sized systems. The systems parameters are $R_V=\SI{2}{\ohm}$, $R_0= R/N =\SI{200}{\ohm}$, $p_\mathrm{err}=0.01$ and $V=\SI{1}{\volt}$.}
\end{figure}

Figure \ref{fig:minimal-gain} shows how the minimal gain $\lambda_\mathrm{min}$
affects the agents' behaviour quantified by average number of cooperators in the system $c_\mathrm{avg}$ for two different system size $N$.
%
%
One can see that bigger values of $\lambda_\mathrm{min}$ lead to more cooperation in the system and different behaviour patterns.
But only in mid-sized systems (as when $N=100$) one can see wave-like patterns for certain ranges.
We can then infer that the system is dominated by the agent behaviours for high and low thresholds of $\lambda_{\mathrm{min}}$, while complex behaviour only emerges in few cases where a proper interplay between the layers occurs.

To have a better understanding on how the minimal gain influences the behaviour, we need to analyse (3).
We find that the upper limit for the tipping  point is given by $\frac{1}{\lambda_{\mathrm{min}}}$ (more details in Section \ref{sec-methods}),  meaning that after such a point the gain of adding another resistor for a single agent becomes too small. 
This stands in contrast to the global optimum, which represents the tipping point after which another resistor added to the system as a whole will result in reduced power delivered to the agents.
Only in the case that all agents behave equally at every point in time, i.e. $a_i[t]=a_\mathrm{avg}[t] \; \forall \; i \in \mathcal{A}$, the tipping point for each agent also becomes the global optimum of $a_{\mathrm{avg}}[t]=\mu$.

As the system grows, the feedback that each agent can infer is then reduced, so is its influence on the system as a whole.
For large systems the deviation of a single agent from the average of the rest of the system has little influence on the average of the whole system.
This reduced feedback  leads to a situation where agents will become trapped in a state of what we call \textit{cooperative solidarity}. 
Agents in this trapped state have reduced their number of resistors to minimum (i.e. only one active resistor), while having still not seen a positive gain.
Consequently, they remain cooperative as do their neighbours.

In this scenario, all resources of the system will be used by the agents that are not yet trapped, leading to high inequality levels. 
For large systems the feedback about individual behaviours are so small that the only way to skip from this solidarity trap is, surprisingly, through communication errors.
In this case, the trapped agents (wrongly) believe that some neighbours stopped cooperating. 

For middle-sized systems, however, the gain may become large enough if a sufficient number of agents synchronise in reducing their number of resistors. 
This results in a positive gain to all the trapped agents since the system recovers from a state of overusing. 
This positive gain then leads to agents that no longer seek cooperation as they are already in the lowest state possible and the system has recovered. 
This in turn explains the wave-like behaviour in mid-size systems.

\begin{figure}[!t]
\centering
\includegraphics[width=1\columnwidth]{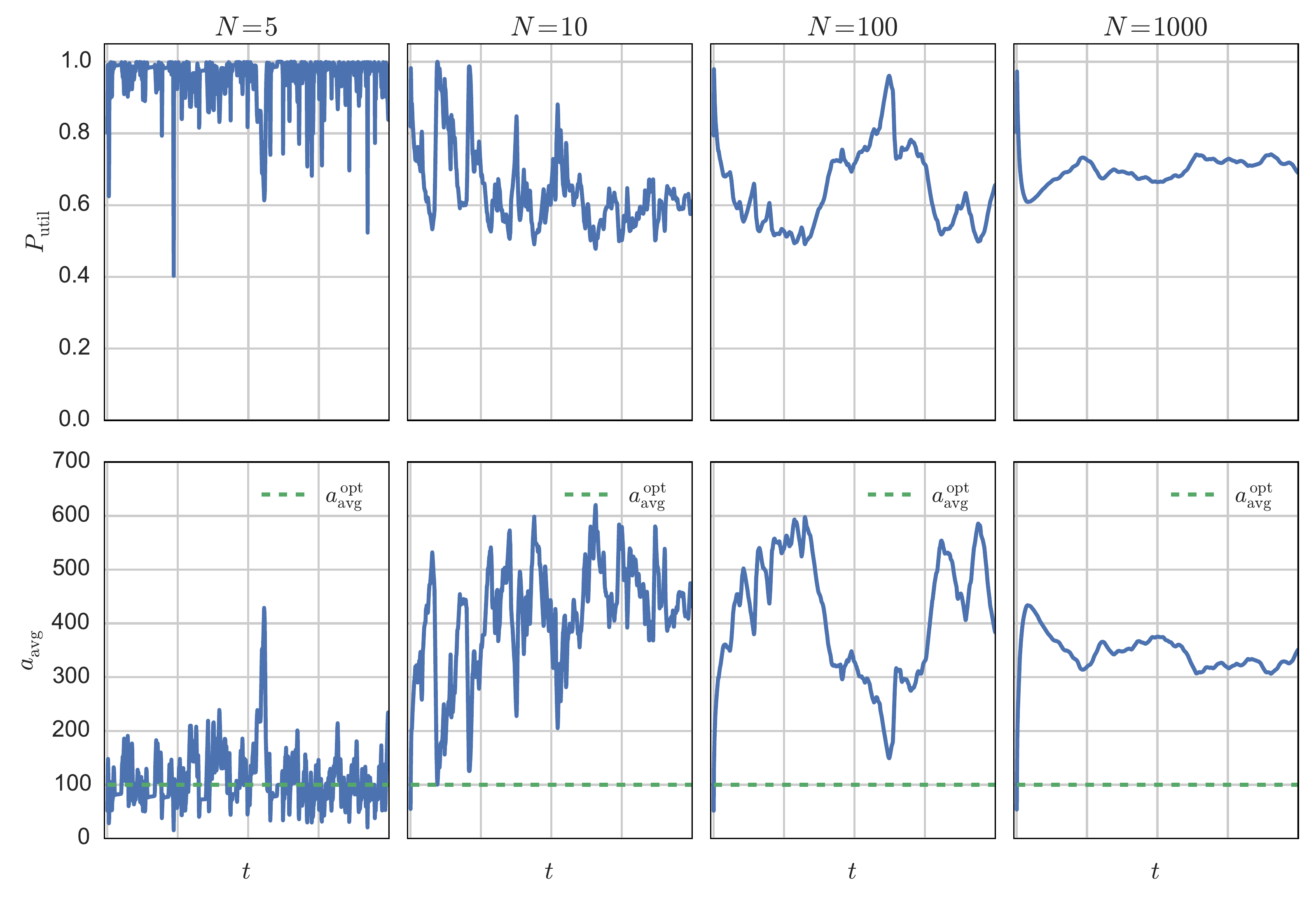}
\caption{ \label{fig:behaviour-plots} Typical behaviour of very small ($N=5$), small ($N=10$), medium ($N=100$) and large systems ($N=1000$). The small systems can operate much closer to the optimum, while mid-size systems show big jumps after stable periods. The system parameters are  $\lambda_\mathrm{min}=0.005$, $R_{\mathrm{V}}=\SI{2}{\ohm}$, $R_0= R/N =\SI{200}{\ohm}$, $p_\mathrm{err}=0.01$ and $V=\SI{1}{\volt}$.}
\end{figure}

Figure \ref{fig:behaviour-plots} shows the system from a different perspective. 
Let us first define the power utilisation $P_{\mathrm{util}}$ as the fraction of power that is utilised by the system and the available power:
\[
	P_{\mathrm{util}}=\frac{4}{P_{\mathrm{typ}}} \;\sum_{i \in \mathcal{A}}P_{i,\mathrm{avg}},
\]
where $P_{i,\mathrm{avg}}$ is the time average power of agent $i$ computed as:
\[
	P_{i,\mathrm{avg}}=\; \lim\limits_{T \rightarrow \infty} \; \dfrac{1}{T} \sum\limits_{t=0}^{T-1} P_i[t].
\]

By looking at the power utilisation $P_{\mathrm{util}}$ and the average number of resistors $a_{\mathrm{avg}}$, one can see that both are related to each other\footnote{Note that all systems in this example are stable in the sense that number of resistors varies within a certain range and does not diverge. This is not necessarily true. Depending on the choices of $\mu$ and $\lambda_{\mathrm{min}}$ the system can behave differently, e.g. all agents remain with only one resistor, or end up collecting more and more resistors.}.
How close the system can operates to the optimum wildly varies between the different system sizes. 
The only system that can operate very close to the optimum is the very small system with a size of only five agents ($N=5$).
We also see the sudden jumps in behaviour for mid-size systems, while large ones appear to be the most stable.

\begin{figure}[!t]
\centering
\includegraphics[width=0.7\columnwidth]{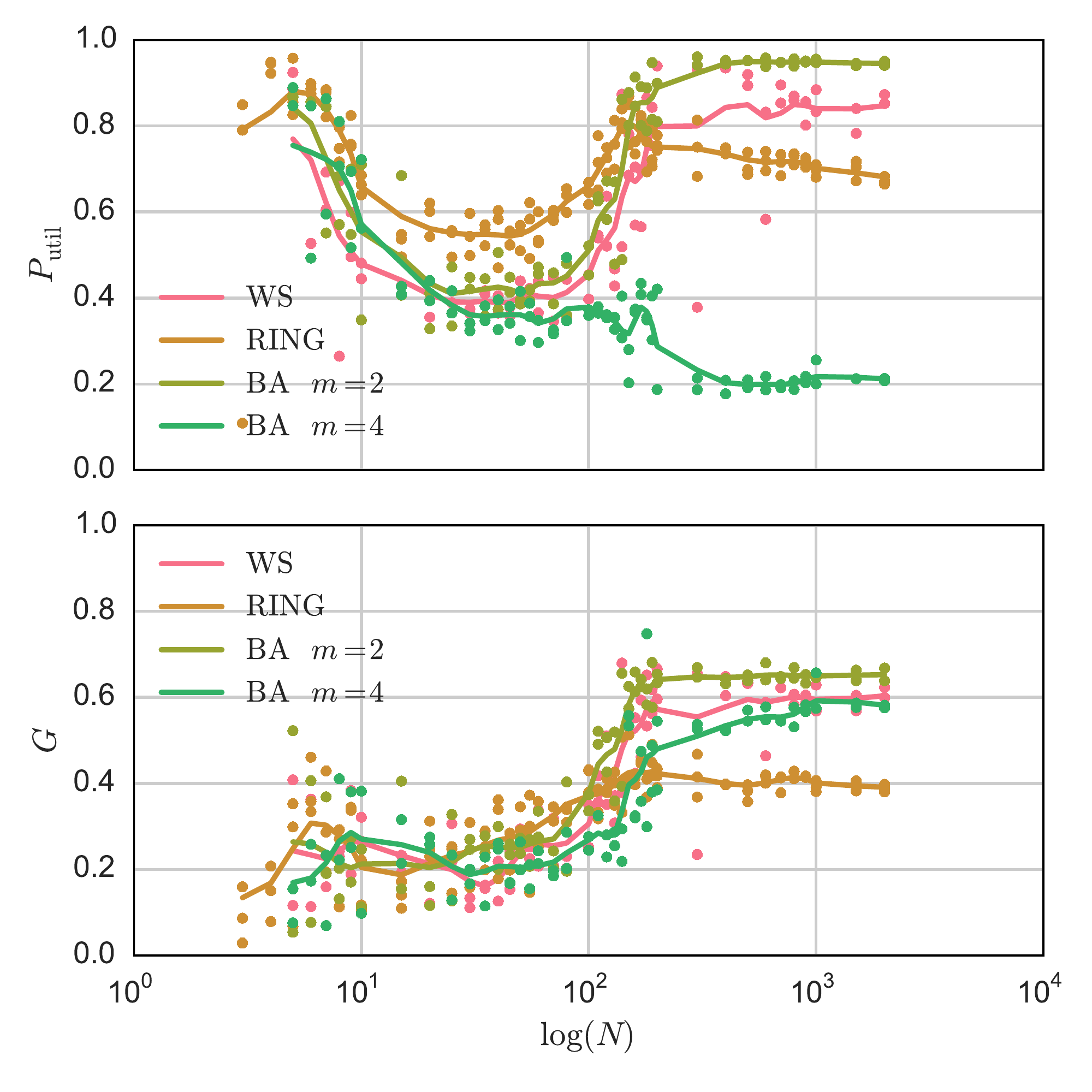}
\caption{\label{fig:Power-utilization-inequality} Power utilisation $P_\mathrm{util}$ (top) and Gini index $G$ (bottom) as a measure of inequality of power distribution, where  $G=1$ representing the biggest inequality. The
Watt-Strogatz-Graph (WS) setting is: mean degree $K=4$, rewiring probability
$\beta=0.5$. Barabasi-Albert-Graph (BA) setting is: number of nodes to attach
to $m=2,4$. The system parameters are: $\lambda_{min}=0.005$, $R_{\mathrm{V}}=\SI{2}{\ohm}$, $R_0= R/N =\SI{200}{\ohm}$, $p_\mathrm{err}=0.01$ and $V=\SI{1}{\volt}$.}
\end{figure}

To better understand what happens in large systems, we analyse Figure \ref{fig:Power-utilization-inequality}, which illustrates the changes in power utilisation and inequality (fairness) in power usage between the agents depending on the size of the system $N$ and different topologies of communication network.
The inequality (fairness) is measured here by the Gini index \cite{Gini:1912}:
\[
G=\dfrac{2}{n} \; \left(\dfrac{\sum\limits_{i\in \mathcal{A}}iP_{i,\mathrm{avg}}}{\sum\limits_{i\in \mathcal{A}}P_{i,\mathrm{avg}}}\right)-\frac{n+1}{n}\text{ with }P_{i,\mathrm{avg}}\leq P_{i+1,\mathrm{avg}},
\]
where total equality is $G=0$  and the biggest inequality is $G=1$.

We see that, for small systems, the results are very close together specially for the complex network Watt-Strogatz and Barabasi-Albert \cite{Newman2006}. 
When the system size increases, on the other hand, one can see a growing difference between the results of different topologies, suggesting that the communication layer plays a big role in the system dynamics.

One can also see a drastic change for the Barabasi-Albert network with $m=4$ that shows a much lower power utilisation than all other networks. 
This is due to the fact that the lowest degree of any node in this network is four.
This means that, for a given agent breaking free from solidarity trap, at least four communication errors must happen, instead of two in the ring topology or only one in Watt-Strogatz (for the nodes with a degree of one). 
Therefore, the power is underutilised for the Barabasi-Albert network. 

The Gini index analysis indicates, that when the system size grows, few agents receive most ot the power, which from a global perspective makes the system very stable.
We also see how the outcome is dominated by the structure of the communication network when large systems are considered.

\begin{figure}[!t]
\centering
\includegraphics[width=1\textwidth]{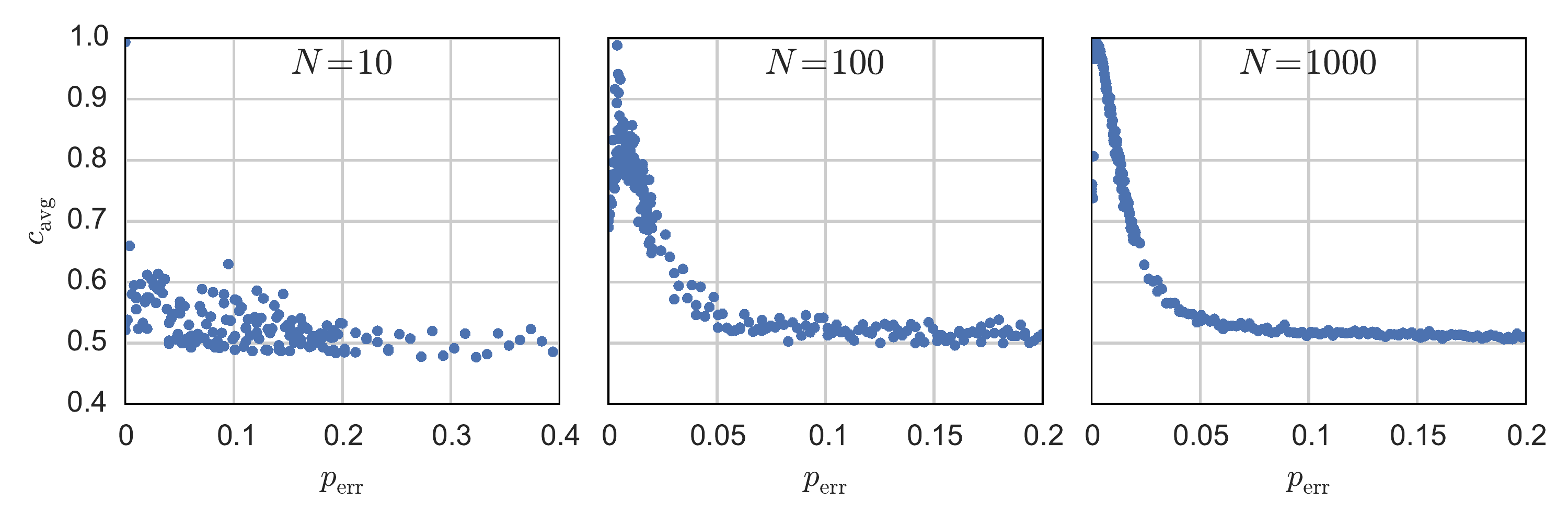}
\caption{\label{fig:Influence-of-communication} Influence of communication error probability $p_\mathrm{err}$ on systems with different sizes $N$. The parameters are $\lambda_\mathrm{min}=0.005$, $R_{\mathrm{V}}=\SI{2}{\ohm}$, $R_0= R/N =\SI{200}{\ohm}$ and $V=\SI{1}{\volt}$.}
\end{figure}

Figure \ref{fig:Influence-of-communication} presents more evidence on how the communication system starts dominating the system dynamics when the system size $N$ grows. 
For small systems, one can only see a small reaction with a lot of scattering when rising the communication error probability.
Surprisingly we can see global maxima appearing for mid-size and large systems. 
This fact may indicate that the communication error probability has a similar effect on our systems as temperature on the susceptibility of physical systems \cite{Wolf:2000}. 
When $p_\mathrm{err} = 0$, the system can become trapped in a state of solidarity with 100\% cooperation (minimal power consumption). 
With increasing the error probability, the system has a random aspect that allows for agents to defect. 
However, if the error probability becomes too high, the state information exchange becomes worthless and the system is dominated by the randomness.

\section{Discussions}
We believe that our proposed multi-layer system can indeed emulate features of a real-world system with coupled physical, communication and decision layers.
Nonetheless, our framework has not been developed to model any specific infra-structure. 
Our idea is to construct a toy-model where the components are rather simple and easy to understand, but where unpredictable behaviour can emerge in certain circumstances, resembling real-world phenomena as modern power grids \cite{nardelli2014models} or cities \cite{batty2013cities}. 
As previously mentioned, large-scale systems built upon those layers are getting more and more usual.
In any case, we believe that it is also worth discussing the design of the components employed herein. 
Let us first deal with the rules the agents follow. 
One might compare the agents to humans or machines acting in behalf of humans. 
In either case, the possible variety of behaviours may be as large as infinite. 
The same can be said for the interactions that occur between the agents.
In this work, the decision procedure and the communication network connections have been arbitrarily chosen to be understandable and justifiable.
Although our simulation assume the network topology and the decision rule (including the selfishness gene and the minimal power gain) as given, both of them could be evolved as part of the simulation.
We could argue that our model allows for complete explanations in the sense that we need not to find explanations that are external to the system; everything can be constructed within the system domain. 
This, however, might mask the results by, for example, slowing down changes in behaviour \cite[Ch.5]{hodgson2012pleasure}.

For the basic agent principle of maximising power, it is important to remember that usually power consumption is just a by-product of making life more comfortable and less manual labour intensive by using more loads in the electricity power grid. 
Similarly to humans, the agents assumed here are most of the time reluctant to remove a resistor once it is installed. 
They only consider to do so when they did not had a large enough gain from adding a resistor (which means that the system might be overused) or when the social pressure is too much (by being non-cooperative in a cooperative neighbourhood).

To further justify the concept of cooperative solidarity, one has to understand the prisoners-dilemma-type of situations \cite{Gianetto2015}.
If the system is close to the optimal point, the agents see a very low gain so that they would have to change their strategy. 
If most of them reduce their number of resistors, they might receive a little less power (in the case of the system being on the left side of the optimum) or they receive a little more by getting closer to the optimum.
However, if only a certain percentage of them reduces their number of resistors while others raise their numbers, the latter gain more power even if the system is very overused. 
If most of them add resistors, however everyone is worse off than before, which resembles the tragedy of the commons \cite{Hardin1968}.

One of the core findings of this report is that, not only is it possible to create systems with complex behaviour through a combination of a few simple parts (as in \cite{Wolfram2002}), but every layer of the system has an influence.
And in some cases, one layer can even dominate the global system behaviour.
A special note should be taken in the strong size-dependency of the model, since as far as real-world engineering systems are always subject to changes in  the number of users after the initial deployment, leading to unforeseen situations (e.g. power grids \cite{nardelli2014models}, cities \cite{batty2013cities} or highways \cite{helbing2001traffic}).

Another interesting aspect can be found in the system behaviour in response to communication errors. 
As was shown in Figure \ref{fig:Influence-of-communication}, a sharp peak exists for the number of cooperators in the system for a communication error probability of about 1\%. 
From an engineering standpoint, one might prefer a system without errors, which (ideally) leads to a stable and a more predictable behaviour. 
However, the existence of even a small amount of errors leads to a significant change in the behaviour and sometimes it might even be desired (e.g. to unfreeze the system from solidarity trap).  

Let us assume that the system should work on a state of very high cooperation.
Our results indicate that an external attacker trying to disturb the system does not have to shut down the whole communication network to break the dominant cooperative state of the whole system.
Rather, it would suffice to generate a small amount of randomly-generated erroneous messages as this will unfreeze the system.
This might impose a new systemic dynamics. 
For an attacker, this would then mean that he does not have to capture the whole communication network to disturb the desired behaviour, due to the coupled nature of the system as a whole.
Consequently, security precautions should be designed accordingly. 

\section{Methods}
\label{sec-methods}

\subsection{Physical layer}
The physical layer used in this paper is depict in Figure \ref{fig:Electrical-System}.
The value of the resistors $R$ that are under the control of the agents $i \in \mathcal{A}$ should be scaled with the number of agents in the system so that
\[
R=NR_0,
\]
where $R_0$ is a constant value, arbitrarily chosen, independent from the system size.

If all agents have in total $k$ appliances, the equivalent resistance of the system
is $R_\mathrm{eq} = \frac{R}{k}$ ($k$ resistors of $R$ in parallel) \cite{dorf2010introduction}. 
The system starts with the different agents having
a random number of resistors so that the equivalent resistance is above
the system-wide optimal point $R_\mathrm{eq}^*=R_{\mathrm{V}}$, computed in terms of power consumed by the agents. 

Let us now describe the system from the point of view of a single
agent $i \in \mathcal{A}$. 
For agent $i$, all the other individual agents can be combined at time-slot $t$
into
\[
R_{\mathrm{sys},i}[t]=\frac{R}{r_i[t]},
\]
where $r_i[t]$ stands for the number of resistors in the system that do
not belong to agent $i$. 
The agent itself is then described by
\[
R_{i}[t]=\frac{R}{a_i[t]}.
\]

We can then derive the power $P_{i}[t]$ that agent $i$ consumes at time-slot $t$ by
\[
P_{i}[t]=\frac{(V\sqrt{N}-IR_{\mathrm{V}})^{2}}{R_{i}[t]}=P_{\mathrm{typ}}\frac{a_i[t]\mu N^{2}}{(a_i[t]+r_i[t]+N\mu)^{2}},
\]
where $V$ is the voltage source, $I$ is the electric current passing through the source resistor $R_\mathrm{V}$, $P_{\mathrm{typ}}=\frac{V^{2}}{R_\mathrm{V}}$ and $\mu=\frac{R}{R_{\mathrm{V}}}$.
Note that we scale the voltage with the square-root of the system size
so that the power available for the agents stays constant.
We could then write $P_{i}$ as
\begin{equation*}
P_{i}[t]=P_{\mathrm{typ}}\frac{a_i[t]\mu}{(a_{\mathrm{avg}}[t]+\mu)^{2}},\label{eq:mean-field}.
\end{equation*}
where $a_{\mathrm{avg}}[t]=(a_i[t]+r_i[t])/N$.

\subsection{Communication layer} 
The communication layer allows for exchange of information between the agents, which is a necessary condition to coordinate their actions in the system. 
In this case, we need to describe what kind of information is sent by the agents and how they build links creating then their neighbourhood set.
Every agent $i \in \mathcal{A}$ sends to their neighbours (to be defined next) a message containing its own state in that time step $t$ such that $S_i[t] \in \{-1,0,+1\}$.
The transmitted message may be detected in error with a given error probability $p_{\mathrm{err}} \in [0,1]$, which is independent from anything else and uniformly distributed.
If an error event in the message detection happens, the receiving agent read one of the two other possible states with the same probability.
In this case, the error probability $p_{\mathrm{err}}$ is the only parameter we have control.
It is worth mentioning that, although we assume $p_{\mathrm{err}}$ as given, it in fact is a result of the communication strategy used \cite{marsic2010computer}.
Nevertheless, we believe that this more realistic approach goes beyond the focus of this work.

The topology of the communication network defines the neighbourhood set of the agents.
In this report we focused on three classes of graphs: ring, Watts-Strogatzs (WS) and Barabasi-Albert (BA) \cite{Newman2006}.
The ring network is defined in a way that agent $i$ is connected with to the agents $i-1$ and $i+1$ (refer to Figure \ref{fig:Electrical-System}).
In this case, agent $1$ is connected to agent $N$ and vice-versa, so the graph topology resembles a ring

To define more complex neighbourhoods, we employ the Watts-Strogatzs (WS) and the Barabasi-Albert (BA) graphs for social networks.
The WS graph is constructed as follows.
The graph starts with a regular lattice where each node has exactly $K$ neighbours.
The links are then rewired with probability $\beta$ resulting in a more random structure.
The BA graph is formed by adding the desired number of nodes step-by-step, starting with a small initial set. 
Each new node is then preferably connected to nodes with an already high degree ($k$) , with probability $p_i$ such that
\[
p_i=\frac{k_i}{\sum_j k_j},
\]
generating then a network whose degree distribution is given by a power law.

\subsection{Agent behaviour}

We explain here the behaviour of the agents and its relation to the physical and communication layers previously defined. 
Let us start in the scenario where none of the neighbours of agent $i$, defined by the neighbourhood set $\mathcal{N}_i$, is cooperative.
This is the default state in the beginning.
Agent $i$ will then behave randomly according to its selfishness gene $s_i$.
The selfishness gene is a random number with uniform distribution assigned to every agent $i \in \mathcal{A}$ before start the system simulation itself. 

The decision procedure is the following: a random number $\xi$ is drawn. 
If $\xi>s_i$,  agent $i$ will switch to the cooperative mode and remove one resistor.
Otherwise, it will switch to one of the non-cooperative modes.
This means that, when $s_i$ is big, agent $i$ is more selfish.
Conversely, agent $i$ is more cooperative when $s_i$ is small

For the non-cooperative modes, the agent will again draw a random number do decide if it will add a resistor if $\xi<s_i$ or do nothing otherwise . 
The decision process is consistent, but depends on random variables. 
Therefore the higher the selfishness, the higher the probability that an agent starts accumulating
resistors.

The agent behaviour also depends on the minimal gain
$\lambda_{\mathrm{min}}$ that the agents needs to stick to their strategy. 
We assume that $\lambda_{\mathrm{min}}$ is fixed and pre-defined before the simulation.
The gain $\lambda_i[t]$ that agent $i$ has at time-slot $t$ is computed as
\[
\lambda_i[t]=\dfrac{P_{i}[t]-P_{i}[t-1]}{P_{i}[t-1]}=\dfrac{\Delta P_{i}[t]}{P_{i}[t-1]}.
\]

Let us now assume that the functions of $t$ are continuous so that  we can use the total derivative rule as follows:
\[
\mathrm{d}P_{i}=P_{typ}\left[\frac{\mu}{(a_{\mathrm{avg}}[t]+\mu)^{2}}\mathrm{d} a_i-\frac{2}{N}\frac{a_i\mu}{(a_{avg}[t]+\mu)^{3}}\mathrm{d}a_i-\frac{2}{N}\frac{a_i\mu}{(a_{\mathrm{avg}}[t]+\mu)^{3}}\mathrm{d}r_i\right].
\]

Now returning to the discrete domain, we consider $\Delta P_i[t] \approx \mathrm{d}P_i$, $\Delta a_i[t] \approx \mathrm{d}a_i$, $\Delta r_i[t] \approx \mathrm{d}r_i$ and $P_i[t] \approx P_i[t-1]$. 
Then, the gain $\lambda_i[t]$ can be evaluated as
\begin{equation}
\lambda_i[t]\approx \frac{\mathrm{d}P_{i}}{P_{i}[t]}\approx\frac{\Delta a_i[t]}{a_i[t]}-\frac{2}{N}\frac{1}{a_{avg}[t]+\mu}\left(\Delta r_i[t]+\Delta a_i[t]\right)=\frac{\Delta a_i[t]}{a_i[t]}-\frac{2}{a_{\mathrm{avg}}[t]+\mu}\Delta a_{\mathrm{avg}}[t].
\end{equation}

The advantage of proceeding in this way is that one can see that the that the gain not only depends on the amount of resistors agent $i$ possesses and the behaviour of the system, but also on the system size $N$.
The higher the number of resistors agent $i$ has, the smaller the first term is since $\Delta a_i[t]$ can only be $-1$, $0$, or $+1$. 
For a very large system ($N\rightarrow \infty$) and agent $i$ non-cooperative (i.e. $\Delta a_i[t]$=1), agent $i$ reaches the minimum gain when it reaches
\[
a_\mathrm{max}=\frac{1}{\lambda_\mathrm{min}}.
\] 

For example, the minimal gain $\lambda_\mathrm{min}=0.0005$ used for most of the simulation scenarios leads to $a_\mathrm{max}=2000$, which is far above the optimal point $a_{\mathrm{avg}}^*=\mu=100$. 

The second term, which has a negative leading sign, provides some sort of feedback.
We could then split the term in the feedback from individual actions $\Delta a_i[t]$ and external actions $\Delta r_i[t]$. 
The negative leading sign means that this term will reduce the gain in case of rising amount of resistors ($\Delta a_i[t]=1$) or will deliver a positive gain in case of cooperation $\Delta a_i[t]=-1$ or keep its state by doing nothing
$\Delta a_i[t]=0$.
In any case, the effect of such feedback will diminish with rising system size $N$.

\subsection{Simulation tools}

This research was carried out using different open source python tools including: NumPy \cite{Numpy}, iPython \cite{PER-GRA:2007}, NetworkX \cite{hagberg-2008-exploring} and Matplotlib \cite{Hunter:2007}. 
We would like to acknowledge the computing facilities of CSC - IT Center for Science Ltd. \cite{CSC}, which were used to run the simulation scenarios.

\section*{Acknowledgements}
This work was funded by the Finnish Academy (Aka) and CNPq-Brazil under the project \textit{A Theory for Sustainable Smart Grids: Combining Communication Theory, Power Systems, Signal Processing and Economics from a Complexity Science Perspective} (SUSTAIN).

\section*{Author contributions}
The authors equally contributed in designing the study scenarios and in the writing process.
FK performed the simulations and created the figures. 
%

\bibliographystyle{IEEEtran}
\bibliography{ref_abbrev}

\end{document}